\documentclass[twocolumn,pra]{revtex4}

\usepackage{graphicx}
\usepackage{dcolumn}
\usepackage{bm}
\usepackage{theorem}

\def\squareforqed{\hbox{\rlap{$\sqcap$}$\sqcup$}}
\def\qed{\ifmmode\squareforqed\else{\unskip\nobreak\hfil
\penalty50\hskip1em\null\nobreak\hfil\squareforqed
\parfillskip=0pt\finalhyphendemerits=0\endgraf}\fi}
\def\endenv{\ifmmode\;\else{\unskip\nobreak\hfil
\penalty50\hskip1em\null\nobreak\hfil\;
\parfillskip=0pt\finalhyphendemerits=0\endgraf}\fi}

\newcommand{\ketbra}[2]{|#1\rangle\!\langle#2|}

\begin{document}
\input epsf
\title{\bf \large Insights into classical irreversible computation using quantum information concepts}

\author{Berry Groisman}
\affiliation{Centre for Quantum Computation, DAMTP, Centre for
Mathematical Sciences, University of Cambridge, Wilberforce Road,
Cambridge CB3 0WA, United Kingdom. }


\begin{abstract}
The method of using concepts and insight from quantum information
theory in order to solve problems in reversible classical computing
(introduced in Ref. \cite{bound_Toffoli}) have been generalized to
irreversible classical computing. The method have been successfully
tested on two computational tasks. Several basic logic gates have
been analyzed and the nonlocal content of the associate quantum
transformations have been calculated. The results provide us with
new interesting insight into the notion of complexity of logic
operations.
\end{abstract}


\maketitle

\section{Introduction}\label{intro}
The field of quantum information theory was born from a symbiosis of
classical information theory and quantum mechanics. The main driving
force behind the development of the new field is a vision of new
computation and communication protocols and devices that would
outperform their classical counterparts. At the same time, one might
expect certain benefits to the classical theory as a ``side effect"
of this development. Such a feedback might be in interesting
manifestation of the dynamical coexistence of the two fields from
both theoretical and practical perspectives.

The first steps in this direction were made in Ref.
\cite{bound_Toffoli}, where a method of evaluating (the bounds on)
the number of Toffoli gates in a classical reversible circuit using
quantum information concepts was proposed.

Here we generalize this approach to irreversible classical circuits,
which provides an interesting insights into the nature of classical
computing.

The structure of the article is as follows. Section \ref{sec:method}
gives a brief overview of the method in \cite{bound_Toffoli}.
Section \ref{sec:generalization2irreversible} presents the details
of how the method is generalized to irreversible case. In Section
\ref{sec:method} we give two examples of its implementation.

\section{The Method}\label{sec:method}
Let us recall the basic principles of the approach in
\cite{bound_Toffoli}. The key idea is to {\it map} classical bits
onto special orthogonal quantum states, i.e. $0 \rightarrow  |{\bf
0}\rangle$ and $1 \rightarrow  |{\bf 1}\rangle$, thereby mapping
strings of $n$ bits on the associate products of $n$ quantum states
\begin{equation}
x_1x_2……x_n \rightarrow  |{\bf x}_1\rangle|{\bf
x}_2\rangle……|\bf{x}_n\rangle\end{equation} such as
\begin{equation}001010\rightarrow |{\bf 0}\rangle|{\bf 0}\rangle|{\bf
1}\rangle|{\bf0}\rangle|{\bf 1}\rangle|{\bf 0}\rangle.\end{equation}
Subsequently, the action of the logic gates is mapped onto
corresponding transformations acting on these states. For reversible
circuits the corresponding associate quantum transformations are
unitary. Then, the study of the properties of the quantum
transformation that is associated to the classical computation can
provide information about the classical circuit. The property of a
quantum transformation was chosen to be its nonlocal content as
expressed via two quantities: first, its entangling capacity
$E^{\uparrow}$, second its entanglement cost, $E_{cost}$ (which
always satisfy $E^{\uparrow}\leq E_{cost}$) \cite{footnote_1}. The
number of elementary gates needed to implement a particular
classical computation is bounded from below by the following ratio
for the associate quantum transformations
\begin{equation}\label{bound}
N_{gates}\geq \frac{E^{\uparrow}(circuit)}{E_{cost}(gate)}.
\end{equation}
To realize this programme the two states $|{\bf 0}\rangle$ and
$|{\bf 1}\rangle$ have to be nonlocal (entangled) states. In
particular, the following states were used in \cite{bound_Toffoli}

\parbox{0.3in}{\begin{eqnarray*}\end{eqnarray*}}\hfill
\parbox{2.in}{\begin{eqnarray*} |{\bf
0}\rangle=\frac{1}{\sqrt{2}}(|0\rangle_A|0\rangle_B+|1\rangle_A|1\rangle_B),
\\ |{\bf
1}\rangle=\frac{1}{\sqrt{2}}(|0\rangle_A|0\rangle_B-|1\rangle_A|1\rangle_B).\label{map}
\end{eqnarray*}}\hfill
\parbox{0.5in}{\begin{eqnarray}\end{eqnarray}}

Thus, classical logic bits are encoded in orthogonal maximally
entangled states of two qubits.

\section{Generalization to irreversible}\label{sec:generalization2irreversible}
 Here we generalize this approach to classical
irreversible circuits. In this case the associate quantum
transformations become non-unitary. In addition to entangling
capacity and entanglement cost we will associate a third quantity -
{\it disentangling capacity}, $E^\downarrow$ - with a non-unitary
operation. For in the case of reversible computing $E^\downarrow$
was not very important since it does not affect the bound in Eq.
(\ref{bound}). For irreversible computing the interplay between
$E^\uparrow$, $E^\downarrow$ and $E_{cost}$ become much more
interesting.

\subsection{1-bit logic gates: RESET}
Let us start with a trivial example of the simple logic operation
RESET (to zero) with corresponding truth-table
\begin{eqnarray}
0\rightarrow   0\nonumber\\
1\rightarrow   0\nonumber
\end{eqnarray}
which is mapped on the non-unitary transformation $G_{reset}$
\begin{eqnarray}
|{\bf 0}\rangle\rightarrow |{\bf 0}\rangle~\nonumber\\
|{\bf 1}\rangle\rightarrow |{\bf 0}\rangle.\nonumber
\end{eqnarray}
The operation $G_{reset}$ is a superoperator or CPTP map, and is
able to create 1 ebit of entanglement, which becomes evident if we
consider unentangled input state $1/\sqrt{2}(|{\bf 0}\rangle+|{\bf
1}\rangle)$. On the other hand 1 ebit is also sufficient to
implement $G_{reset}$. This task can be accomplished, for example,
if one simply replaces the original pair by an ancillary pair in a
state $|{\bf 0}\rangle$ and discards the original pair. More
formally, this procedure can be presented as a SWAP operation on the
original pair and an ancillary pair in the state $|{\bf 0}\rangle$
followed by discarding the ancillary pair. Thus,
$E^{\uparrow}(G_{RESET})=E_{cost}(G_{RESET})=1$. It is obvious that
$G_{RESET}$ cannot destroy any entanglement since it always has a
maximally entangled state as its output, i.e.
$E^{\downarrow}(G_{RESET})=0$.

Corollary, $E^{\downarrow}=0$ for a quantum counterpart of any
deterministic (surjective) logic operation, with equal numbers of
inputs and outputs.

\subsection{2-bit logic gates: XOR}
Let us analyze a more complex, two-bit, XOR gate
\begin{eqnarray}
00\rightarrow   0\nonumber~\\
01\rightarrow   1\nonumber~\\
10\rightarrow   1\nonumber~\\
11\rightarrow   0\nonumber.
\end{eqnarray}
It is mapped on the non-unitary transformation $G_{XOR}$
\begin{eqnarray}
|{\bf 0}\rangle|{\bf 0}\rangle\rightarrow |{\bf 0}\rangle~\nonumber\\
|{\bf 0}\rangle|{\bf 1}\rangle\rightarrow |{\bf 1}\rangle~\nonumber\\
|{\bf 1}\rangle|{\bf 0}\rangle\rightarrow |{\bf 1}\rangle~\nonumber\\
|{\bf 1}\rangle|{\bf 1}\rangle\rightarrow |{\bf 0}\rangle.\nonumber
\end{eqnarray}
Note, that unlike 1-bit transformations, $G_{XOR}$ is not trace
preserving, because it is accompanied by loss of subsystems. We can
account for this loss by introducing a purification. For example,
$G_{XOR}$ can be obtained if one implements (unitary) CNOT
\begin{eqnarray}
|{\bf 0}\rangle|{\bf 0}\rangle\rightarrow |{\bf 0}\rangle|{\bf 0}\rangle~\nonumber\\
|{\bf 0}\rangle|{\bf 1}\rangle\rightarrow |{\bf 0}\rangle|{\bf 1}\rangle~\nonumber\\
|{\bf 1}\rangle|{\bf 0}\rangle\rightarrow |{\bf 1}\rangle|{\bf 1}\rangle~\nonumber\\
|{\bf 1}\rangle|{\bf 1}\rangle\rightarrow |{\bf 1}\rangle|{\bf
0}\rangle.\nonumber
\end{eqnarray}
and discards the first pair. Nonlocal CNOT transformation is
equivalent to two local CNOT transformations on corresponding qubits
and therefore has zero entanglement cost \cite{bound_Toffoli}.
Therefore, $E_{cost}(G_{XOR})=0$. However, since the first pair is
discarder it effectively involves ``dissipation" of entanglement. In
other words, whatever the mechanism inside the box is it will
``dissipate" entanglement of the first pair. Therefore,
$E^{\downarrow}(G_{XOR})=1$, $E^{\uparrow}(G_{XOR})=0$.

\subsection{Universal 2-bit gates: NAND and NOR}
Conceptually, the most interesting are NAND and NOR gates, because
each of them is a universal gate for irreversible computation. Let
us consider now the NAND gate
\begin{eqnarray}
00\rightarrow   1\nonumber~\\
01\rightarrow   1\nonumber~\\
10\rightarrow   1\nonumber~\\
11\rightarrow   0\nonumber.
\end{eqnarray}
It is mapped on the non-unitary transformation $G_{NAND}$
\begin{eqnarray}\label{G_AND}
|{\bf 0}\rangle|{\bf 0}\rangle\rightarrow |{\bf 1}\rangle~\nonumber\\
|{\bf 0}\rangle|{\bf 1}\rangle\rightarrow |{\bf 1}\rangle~\\
|{\bf 1}\rangle|{\bf 0}\rangle\rightarrow |{\bf 1}\rangle~\nonumber\\
|{\bf 1}\rangle|{\bf 1}\rangle\rightarrow |{\bf 0}\rangle.\nonumber
\end{eqnarray}
A purification of $G_{NAND}$ may arise if the two original pairs
interact with a third, ancillary, pair in a standard state $|{\bf
0}\rangle$ via unitary transformation
\begin{eqnarray}\label{G_NAND_pur1}
|{\bf 0}\rangle|{\bf 0}\rangle|{\bf 0}\rangle\rightarrow |{\bf 1}\rangle|{\bf 0}\rangle|{\bf 0}\rangle~\nonumber\\
|{\bf 0}\rangle|{\bf 1}\rangle|{\bf 0}\rangle\rightarrow |{\bf 1}\rangle|{\bf 0}\rangle|{\bf 1}\rangle~\\
|{\bf 1}\rangle|{\bf 0}\rangle|{\bf 0}\rangle\rightarrow |{\bf 1}\rangle|{\bf 1}\rangle|{\bf 0}\rangle~\nonumber\\
|{\bf 1}\rangle|{\bf 1}\rangle|{\bf 0}\rangle\rightarrow |{\bf
0}\rangle|{\bf 1}\rangle|{\bf 1}\rangle,\nonumber
\end{eqnarray}
and consequently the ancilla and the second pair are discarded
(traced out). In other words, it is a trace out of a Toffoli gate
with one standard input. It should be noted that the realization
(purification) in Eq. (\ref{G_NAND_pur1}) is not unique. In fact,
specification of the mapping of basis states in Eq. (\ref{G_AND})
does not fully determine $G_{NAND}$, that is to say that Eq.
(\ref{G_AND}) does not prescribe how an arbitrary linear combination
of basis states will evolve. (This is the nature of a
trace-non-preserving operations - they do not supply full
information about the evolution of the system.) We will come back to
this problem later.

Now let us calculate (bounds on) $E^{\uparrow}$ and $E_{cost}$. It
is a difficult task to calculate the explicit value of
$E^{\uparrow}$ for a quantum operation. At the time this paper was
written the only known method of calculating it was a direct
numerical optimization over all states accessible to the operation
(even including ancillas). The value of $E_{cost}$ can be obtained
by providing an explicit way of implementation, though its
optimality has to be also proven. However, it is much easier to find
bound on these quantities - a lower bound on $E^{\uparrow}$ and an
upper bound on $E_{cost}$. If the two bounds happen to coincide then
one is lucky to obtain exact values of $E^{\uparrow}$ and
$E_{cost}$. As we will see below, our calculations leave the gap
between the two bounds in the case of NAND. Nevertheless, it will
suffice to bound the number of gates $N_{gates}$ as in Eq.
(\ref{bound}).

 To obtain a lower bound on $E^{\uparrow}(G_{NAND})$ we choose a special input state,
 the test state $|\Psi_{in}^{test}\rangle$ and obtain the corresponding output state
\begin{equation}
\rho_{out}^{test}=G_{NAND}(|\Psi_{in}^{test}\rangle).
\end{equation}
The difference between the amounts of entanglement possessed by
$\rho_{out}^{test}$ and $|\Psi_{in}^{test}\rangle$ bounds from below
the entangling capacity of $G_{NAND}$. Any test state that leads to
an increase of entanglement gives a lower bound. The higher the
bound the better. The higher bounds can be obtained either by direct
numerical search of by trial and error.

Consider, for example, the following disentangled state as an input,
\begin{eqnarray}\label{in}
|\Psi_{in}^{test}\rangle={1 \over2}\biggl(|{\bf 0}\rangle_1|{\bf
0}\rangle_2+|{\bf 0}\rangle_1|{\bf 1}\rangle_2+|{\bf
1}\rangle_1|{\bf 0}\rangle_2+|{\bf 1}\rangle_1|{\bf
1}\rangle_2\biggr).
\end{eqnarray}
Of course, the input state for a gate inside an irreversible circuit
will be most likely a mixed state. The choice of the state in Eq.
(\ref{in}) was not motivated by this kind of considerations. We are
interested in lower bounds on the entangling capacity, attainable in
principle. How do we determine $\rho_{out}^{test}$? We have already
mentioned that Eq. (\ref{G_AND}) does not specify $G_{NAND}$
completely. For example, if we extend $G_{NAND}$ to unitary as in
Eq. (\ref{G_NAND_pur1}) then the corresponding output state will be
\begin{equation}
\rho_{out}^{test}={1\over 4}\ketbra{\bf 0}{\bf 0}+ {3\over
4}\ketbra{\bf 1}{\bf 1}.\end{equation} However, the following
unitary
\begin{eqnarray}\label{G_NAND_pur2}
|{\bf 0}\rangle|{\bf 0}\rangle|{\bf 0}\rangle\rightarrow |{\bf 1}\rangle|{\bf 0}\rangle|{\bf 0}\rangle~\nonumber\\
|{\bf 0}\rangle|{\bf 1}\rangle|{\bf 0}\rangle\rightarrow |{\bf 1}\rangle|{\bf 0}\rangle|{\bf 1}\rangle~\\
|{\bf 1}\rangle|{\bf 0}\rangle|{\bf 0}\rangle\rightarrow |{\bf 1}\rangle|{\bf 1}\rangle|{\bf 0}\rangle~\nonumber\\
|{\bf 1}\rangle|{\bf 1}\rangle|{\bf 0}\rangle\rightarrow |{\bf
0}\rangle|{\bf 0}\rangle|{\bf 0}\rangle,\nonumber
\end{eqnarray}
will do the job as well, in which case
\begin{equation}
\rho_{out}^{test}={1\over 4}\ketbra{\bf 0}{\bf 0}+ {3\over
4}\ketbra{\bf 1}{\bf 1}+{1\over 4}\ketbra{\bf 0}{\bf 1}+{1\over
4}\ketbra{\bf 1}{\bf 0}.\end{equation} Thus, $G_{NAND}$ represents
the whole family of quantum maps. The most general unitary extension
will give
\begin{eqnarray}\label{G_NAND_pur3}
|{\bf 1}\rangle|{\bf 1}\rangle|{\bf 0}\rangle\rightarrow |{\bf
0}\rangle(a|{\bf 0}\rangle|{\bf 0}\rangle+b|{\bf 0}\rangle|{\bf
1}\rangle+c|{\bf 1}\rangle|{\bf 0}\rangle+d|{\bf 1}\rangle|{\bf
1}\rangle),\nonumber
\end{eqnarray}
where $a$, $b$, $c$ and $d$ are complex amplitudes. The output
state, therefore, will be
\begin{eqnarray}
\rho_{out}^{test}={1\over 4}\ketbra{\bf 0}{\bf 0}+ {3\over
4}\ketbra{\bf 1}{\bf 1}~~~~~~~~~~~~~~~~~~~~~~\\+{(a+b+c)\over
4}\ketbra{\bf 0}{\bf 1}+{(a^*+b^*+c^*)\over 4}\ketbra{\bf 1}{\bf
0}.\end{eqnarray} It is straightforward to calculate the relative
entropy of entanglement of $\rho$ \cite{VP:ent_m_pp}, which is
\begin{equation}
H[{1\over2}+{1\over4}Re(a+b+c)^2]-H[{1\over2}+{1\over4}\sqrt{1+|a+b+c|^2}].
\end{equation}
It can be shown that the minimum is achieved at $|d|=1$, i.e.
$Re(a+b+c)=Im(a+b+c)=0$ and equal $1-H[1/4]$. Recall that
$E_{in}^{test}=0$ ebits and we obtain $E^{\uparrow}(G_{NAND})\geq
1-H[1/4]=0.189> 0$. On the other hand, it is clear that
$E^{\downarrow}(G_{NAND})=1$

To provide an upper bound on $E_{cost}$ let us consider the
following protocol. In addition to the first two pairs $|{\bf
x_1}\rangle$ and $|{\bf x_2}\rangle$ we use an ancillary pair in a
standard state $|{\bf 0}_3\rangle$ and apply on these three pairs a
nonlocal Toffoli gate as in \cite{bound_Toffoli}. This utilizes 2
ebits. Then we discard first two pairs and are left with one pair in
the desired final state. The total cost of this procedure is 3
ebits. Therefore, $E_{cost}\leq 3$.

Thus, we obtain
\begin{equation}
0.189\leq E^{\uparrow}(G_{NAND})\leq E_{cost}(G_{NAND})\leq 3.
\end{equation}

Interestingly, the NOR gate, an alternative universal gate for
irreversible computing, has the same values of $E^{\uparrow}$,
$E^{\downarrow}$, and $E_{cost}$.

\subsection{Calculating bounds}\label{sec:examples}
Now, if we are given a classical circuit all we have to do is to
calculate (a lower bound on) $E^{\uparrow}$ of the associate quantum
transformation and obtain the bound on $N_{gates}$ using Eq.
(\ref{bound}). As NAND (or NOR) is a universal gate any circuit can
be build from NAND gates alone. The present method is unable to
provide a bound in this case. This is because some of the NAND gates
in a circuit will replace other gates that have $E_{cost}= 0$, and
thus even if the whole quantum transformation has $E^{\uparrow}=0$
we will need nonzero number of NAND gates. On the other hand, we
might consider universal sets of two or more gates, e.g. NAND and
XOR. The bound (\ref{bound}) provides us with a meaningful result
when one tries to minimize the usage of NAND gates by supplementing
them by other
gates, e.g. XOR gates. 

Entanglement cost of the gates in a sense gives us an insight into
the computation complexity of the gates. The fact that
$E_{cost}(G_{XOR})=0$ while $E_{cost}(G_{NAND})>0$ emphasized an
essential difference between XOR and NAND gates - XOR gates involve
less computational effort than NAND gates so to speak.

\section{Example of irreversible computation}
\subsection{Example I: Parity calculation} Consider a trivial example of the circuit that calculates
a parity function, i.e. for a binary string of $n$ bits it
calculates whether the number of zeros is even or odd. Obviously,
half of all possible input strings will have even number of zeros
and half will have odd number of zeroes. Therefore, if we use a
uniform superposition of corresponding quantum sequences as an
input, then the output will be $\rho_{out}={1\over2}|{\bf
0}\rangle\!\langle{\bf 0}|+{1\over2}|{\bf 1}\rangle\!\langle{\bf
1}|$ which has zero entanglement. In fact, numerical calculation
indicate that any input states will lead to no increase of
entanglement, i.e. $E^{\uparrow}=0$. Thus, application of our
method, i.e. Eq. (\ref{bound}), yields $N_{NAND}\geq 0$. Indeed, the
whole computation can be easily accomplished by sequential
application of XOR gates, and therefore no NAND gates are needed to
implement the computation \footnote{The simplest circuit can be
build as a cascade of XOR gates. For odd $n$ one ancillary input
with constant value $x_{n+1}=1$ should be added.}. As we have shown
XOR gates have zero entanglement cost and therefore their
combination cannot increase overall entanglement.

\subsection{Example II: First step in Shannon compression}
As a more complex example let us consider the first step in Shannon
compression - the majority measurement. For $n$ bit input string the
circuit calculates the number of 1's, $k$ \footnote{For $n=3$ this
computation is the well known Full Adder.}. There are $n+1$ possible
values of $k$, therefore we have $m=\lceil\log_2 (n+1)\rceil$ bit
output string
\begin{equation}
x_1x_2...x_n \rightarrow f_1f_2...f_m.
\end{equation}
Without loss of generality let us assume that the output strings
represent the values of $k$ of corresponding input string in a
binary form, e.g.
\begin{eqnarray}
0011\rightarrow10,\nonumber\\
0101\rightarrow10,\\
0111\rightarrow11,\nonumber
\end{eqnarray}
etc.

As before we take the initial state to be a uniform superposition of
all basis states
\begin{equation}\label{Shannon_input}
|\Psi_{in}\rangle={1\over \sqrt{2^n}}\sum_{x_i\in\{0,1\}} |{\bf
x_1}\rangle|{\bf x_2}\rangle...|{\bf x_n}\rangle.
\end{equation}
\begin{figure}
\epsfxsize=3.4truein \centerline{\epsffile{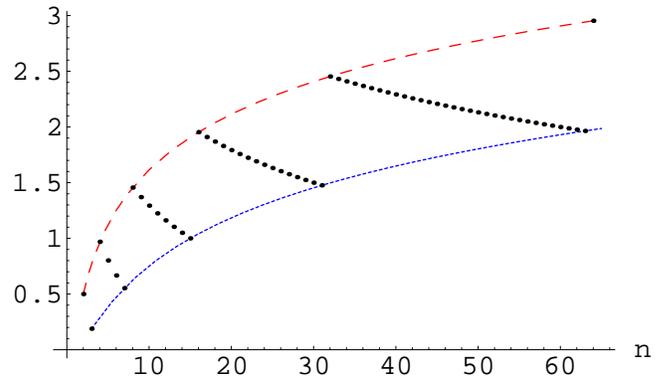}}
\caption[]{The lower bound on the gain of entanglement, as expressed
by $S(Tr_B\rho_{out})-S(\rho_{out})$, for several values of n,
approximated by $0.7055\log_2n -0.0007$ (red dashed line) and
$0.5885\log_2n -0.5324$ (blue dotted line).} \label{fig2}
\end{figure}
The corresponding output state is
\begin{equation}\label{Shannon_output}
\rho_{out}=2^{-n}\sum_{f_i \in \{0,1\}} {n \choose k}\ketbra{\bf
f_1}{\bf f_1}\ketbra{\bf f_2}{\bf f_2}...\ketbra{\bf f_m}{\bf f_m},
\end{equation}
where $k=\sum_{i=1}^{\log_2 n} f_i 2^{i-1}$ is a binary to decimal
converter, that assigns a corresponding value of $k$ to every
sequence (term) in the sum in Eq. (\ref{Shannon_output}). As there
are $\lceil\log_2 (n+1)\rceil$ pairs of qubits involved in the
output state it might give us a clue that the output entanglement
scales not faster than $\log_2 n$. This intuition is supported by
the results showed in Fig. \ref{fig2}.
Therefore, the lower bound on the number of NAND gates required to
implement the computation grows as $0.1962\log_2n-0.1775$. Thus, our
results clearly indicate, that for $n<70$ at least one NAND gate is
required. Interestingly, this conclusion is corroborated by known
circuits implementing the Full Adder ($n=3$), which is constructed
out of two NAND gates, three XOR gates and two NOT gates.

\section{Discussion}
This work demonstrates that the method of Ref. \cite{bound_Toffoli}
can be successfully to irreversible classical computing. Following
that method we have constructed quantum counterparts of several
logic gates, namely RESET, XOR, NAND and NOR, and have calculated
nonlocal content of these quantum transformations. Three quantities
were associated with the entanglement content - entangling capacity,
$E^{\uparrow}$, disentangling capacity, $E^{\downarrow}$ and
entanglement cost, $E_{cost}$. Together with the number of logic
bits these quantities provide a non-trivial characterization of
gates. We have seen that deterministic, but irreversible,
one-to-one-bit gate, RESET, has $E^{\downarrow}=0$ and
$E^{\uparrow}=E_{cost}=1$. An irreversible two-to-one-bit gate, XOR,
exhibits completely different properties, namely $E^{\downarrow}=1$
and $E^{\uparrow}=E_{cost}=0$. The third type - two-to-two
irreversible {\it universal} gates, NAND and NOR, have
$E^{\downarrow}=1$ and $0.189\leq E^{\uparrow}\leq E_{cost}\leq 3$.
The entanglement cost, $E_{cost}$, was an essential ingredient for
estimating lower bounds on the number of universal gates. Every
circuit can be built from the universal gates alone. However, since
NOT and XOR gates have $E_{cost}=0$, our method can only provide
bounds on the minimal number of universal gates when the numbers of
NOT and XOR gates are not limited, i.e. when one aims to minimize
the number of universal gates by using other gates, if possible. It
is worth emphasizing that our method is not constructive in the
sense that the bounds it provides do not tell us how a particular
circuit can be built from a certain type of gates.

The method was tested on two computational task. As a first task we
have chosen the calculation of the parity of a binary string of $n$
bits. Our method indicates that no NAND or NOR gates are needed to
construct a circuit which realizes this calculation. This conclusion
is consistent with the fact, that this computation can be achieved
by simple cascade of XOR gates. The second task is the first step in
Shannon data compression, i.e. calculating the number of $1$s in a
binary $n$-bit string. Our method provides the lower bound on
$N_{NAND}$ of a fraction of $\log_2n$. It is worth noting that our
method can provide minimal bounds of at most $m=\lceil\log_2
(n+1)\rceil$, because the corresponding output quantum state can
posses at most $m$ ebits of entanglement.

Our results provide an interesting insight into complexity of
computation. The link between our approach and computation
complexity was already established in \cite{bound_Toffoli}. Indeed,
in the reversible scenario it was shown that $E_{cost}$ of the
quantum counterpart of Toffoli gate (a universal gate) is non-zero
whereas two-bit gates, e.g. CNOT, and one-bit gates, NOT, do not
consume any entanglement at all. Therefore, one can use $E_{cost}$
as a kind of measure of the complexity of the associate classical
logic gate itself. However, in the reversible case the distinction
between Toffoli and two- or one-bit gates in term of their
complexity is intuitively straightforward - Toffoli is more complex
because it performs a nontrivial computation on more bits. In the
case of irreversible computation this distinction becomes much more
subtle. This is why the results of this work provide us with much
more interesting and powerful insight. Indeed, here universal gates
are not distinguishable from other two-to-one-bit gates simply by
the number of inputs/outputs. than NAND and NOR are in certain sense
computationally (logically) stronger, than XOR. One-to-one-bit RESET
gate stands alone for having a nonzero $E_{cost}$. However, we note
that RESET gate does not actually perform a computational task.

\begin{acknowledgments}
This work was funded by the U.K. Engineering and Physical Sciences
Research Council, Grant No. EP/C528042/1.
\end{acknowledgments}

\end{document}